\documentclass[12pt]{article}
\newcommand{\1}{\begin{equation}}
\newcommand{\2}{\end{equation}}
\newcommand{\ea}{\begin{eqnarray}}
\newcommand{\ee}{\end{eqnarray}}
\newcommand{\bee}{\begin{eqnarray*}}
\newcommand{\eee}{\end{eqnarray*}}

\newcommand{\op}[1]{\hat{#1}}
\newcommand{\erw}[1]{\left\langle\, #1\,\right\rangle}

\newcommand{\de}{{\!\rm d}}
\newcommand{\e}{{\rm e}}

\newcommand{\g}{{\!\,=\,\!}}

\newcommand{\ii}{{\rm i}}

\newcommand{\sa}{\left[ \begin{array} {c} }
\newcommand{\se}{\end{array}\right]}

\usepackage{texdraw}
\usepackage{graphicx}
\begin{document}

\begin{center}
{\Large\bf Analytical pair correlations in ideal quantum gases: Temperature-dependent bunching and antibunching}
\\*[4mm] {\large J. Bosse$^1$}, {\large K. N. Pathak$^2$}, and {\large G. S. Singh$^{3,*}$}\\*[1.5mm]
{\small $^1$Institute of Theoretical Physics, Freie Universit{\"a}t, Berlin 14195, Germany}\\
{\small  $^2$Physics Department, Panjab University, Chandigarh 160 014, India}\\
{\small $^3$Physics Department, Indian Institute of Technology Roorkee, Roorkee 247 667, India\\*[3mm]
(23 October 2011)}\\
*[1cm]

\end{center}

\begin{abstract}
The fluctuation--dissipation theorem together with the exact density response spectrum for ideal quantum gases has been utilized to yield a new expression for the static structure factor, which we use to derive exact analytical expressions for the temperature--dependent pair distribution function $g(r)$ of the ideal gases. The plots of bosonic  and fermionic $g(r)$ display ``Bose pile'' and ``Fermi hole'' typically akin to bunching and antibunching as observed experimentally for ultracold atomic gases. The behavior of spin--scaled pair correlation for fermions is almost featureless but bosons show a rich structure including long--range correlations near $T_c$. The coherent state at $T\g0$ shows no correlation at all, just like single-mode lasers. The depicted decreasing trend in correlation with decrease in temperature for $T<T_c$ should be observable in accurate experiments.
\\*[3mm]

\end{abstract}

\noindent \emph{Introduction}: The surge in the study of various properties of ultracold atomic gases has prompted search for atomic analog of the Hanbury-Brown--Twiss  (HBT) effect \cite{hat:56} by various research groups as reported in Refs. \cite{san:10, mue:10, jel:07} and references therein. The suppression of density fluctuations, a signature of the Pauli exclusion principle at work in real space and thereby antibunching, has been demonstrated in \cite{san:10, mue:10}. On the other hand,  Jeltes \emph{et al.} \cite{jel:07} have compared results of the two-particle correlations of a polarized, but not Bose-condensed, sample of ultracold $^4$He* atoms with those of polarized $^3$He* atoms. The experimental conditions in \cite{jel:07} were such that the gases could be treated almost ideal. Hence bunching for bosons and antibunching for fermions at small interatomic separations have been attributed to purely quantum effects associated with the exchange symmetries of wavefunctions of indistinguishable particles. Also, the measurement of correlations has been reported \cite{sch:05} both above and below the Bose-Einstein condensation (BEC) temperature in atomic $^4$He*.

An elegant form for the correlation function $\nu(r)$ of the density fluctuations in ideal quantum gases (IQG) has been derived by Landau and Lifshitz \cite{lal:93}. Although many qualitative features and limiting expressions for $\nu(r)$, which is related to the pair distribution function $g(r)$, have been discussed by them, quantitative descriptions would require availability of general analytical forms. References \cite{lur:68, gub:68,  bae:71} have discussed expressions for  $g^{BE}(r)$ of an ideal Bose gas (IBG) in one or the other range in the temperature domain $0\!<\!T\!<\!T_c^+$ whereas Lee and Long \cite{lel:95} have given $g^{FD}(r)$ of an ideal electron gas at $T\g0$. Reference \cite{gsb:00} has discussed model analytic expression for the unpolarized homogeneous electron gas in solids. However, these expressions cannot be utilized to get comprehensive theoretical values to compare with the observed HBT effect reported in \cite{jel:07} and \cite{sch:05}. The main purpose of this Brief Report is to fill up the gap by deriving  exact analytical expressions for $g(r)$ of IQGs.  The improvements on some asymptotic results, both for fermions and bosons, available in the literature are also discussed.

A unified approach is presented for evaluation of temperature--dependent $g(r)$ for ideal Bose--Einstein (BE), Fermi--Dirac (FD) and Maxwell--Boltzmann (MB) gases wherein the unification has been achieved using polylogarithm \cite{lew:81, lee:95, g(r):11,lek-lee, bps:10b, bps:10a}. Our method starts from the expression for the density response function derived in \cite{bps:10a} using the method of second quantization, applies the fluctuation--dissipation theorem to get a general new expression for the static structure factor, introduces a function which is related to the one-particle density matrix, and ultimately gets the general form for $g(r)$  valid for all quantum gases at all temperatures.  The computed values of $g^{BE}(r)$ and $g^{FD}(r)$ using our analytical expressions are depicted graphically as a function of $r$ at various temperatures showing at small--$r$ appearance of bump and dip which we term as ``Bose pile'' and ``Fermi hole''. The plots are further compared with the experimental results \cite{jel:07, sch:05} for ultracold atomic gases.

\noindent \emph{Basic Expressions}:
The pair--distribution function $g(r)$ of a uniform
one--component fluid consisting of $N$ particles in volume $V$ is
defined by the thermal average of an operator that counts pairs of
particles located distance $r$ apart, divided by the square of number density.
It is related to the static structure
factor $S(q)$ of the fluid by the spatial Fourier transform:
\1
\label{gofrfromsofq}
n\left[g(r)-1\right]=\frac{1}{V}\sum_{\bf
q}\e^{\ii{\bf q}\cdot{\bf r}}~\left[S(q)-1\right]
\2
with $n=N/V$ denoting the number density. Taking due account of the fact that the operator of the total particle--number, $\op N\equiv N_{\bf q\g0}$, is a   constant of motion, the fluctuation--dissipation theorem for a uniform system, $\chi''(q,\omega)=(n\pi/\hbar) (1-\e^{-\beta\hbar\omega})S(q,\omega)$, can be solved for the van Hove function
$
S(q,\omega)=(\hbar/n\pi)\left(1-\delta_{{\bf q,0}}\right)\,\chi''(q,\omega)/\left(1-\e^{-\beta\hbar\omega}\right)+\erw{(\delta \op N)^2/N}\,\delta_{{\bf q},{\bf 0}}\,\delta(\omega)$, where $\erw{...}$ represents averaging in the grand canonical ensemble (GCE). The expression $S(q)=\int_{-\infty}^\infty\de\omega~S(q,\omega)$ then yields
\1
\label{zero-frequency modes}
S(q)=\left(1-\delta_{{\bf q,0}}\right)\frac{\hbar}{n\pi}\int_0^\infty\de\omega~
\coth\left(\frac{\beta \hbar\omega}{2}\right)\,\chi''(q,\omega) +\delta_{{\bf q,0}}\frac{\erw{(\delta \op N)^2}}{N}.
\2

Upon inserting Eq. (\ref{zero-frequency modes}) in Eq. (\ref{gofrfromsofq}), the ${\bf q}$--sum on the right--hand side separates into two parts: the first part contains the summation with the restriction ${\bf q}\ne{\bf 0}$ while the second results in an additive constant, $\left[n\erw{(\delta \op N/N)^2}-1/V\right],$ which vanishes in the thermodynamic limit for all IQGs except for an IBG at $T\le T_c$. The pathological aspect of GCE for the condensate fluctuations has been ameliorated by replacing $\erw{[\delta(a^\dagger_0a_0)]^2}_{\rm GCE}$ by $\erw{[\delta(a^\dagger_0a_0)]^2}_{\rm CE}$, with $a_0$ and $a^\dagger_0$ being the ground--state annihilation and creation operators, as suggested in Ref. \cite{nag:99} based on results in \cite{pol:96}, and which has been utilized by others, see, e.g., \cite{gom:06}.
The ``law of large numbers''  considered by us in order to make the constant to vanish in the  Bose-condensed phase has the form $\sqrt{\erw{(\delta \op N/N)^2}}\propto N^{-1/3}$, see, e.g., \cite[Eqs. (3.55) \& (3.57)]{pis:03}.

We substitute
$\chi''(q,\omega)=n\pi\sum_{\bf k}C_k\left[\delta\left(\hbar\omega-\Delta_{\bf k}({\bf
q})\right)-\delta\left(\hbar\omega+\Delta_{\bf k}({\bf
q})\right)\right]$ from \cite{bps:10a} for the
IQG with $\Delta_{\bf k}({\bf q})=\varepsilon_{|{\bf k}+{\bf q}|}-\varepsilon_k$, and obtain for ${\bf q}\ne{\bf 0}$\,,
\1
\label{sofq}
S(q)=\sum_{\bf k}C_k\coth\left[\frac{1}{2}\beta\left(\varepsilon_{|{\bf k} + {\bf q}|}-\varepsilon_k\right)\right]; ~~~(q>0)\,,
\2
where $C_k$ denotes the thermal--average fraction of particles having momentum
$\hbar{\bf k}$,
\1
\label{C_k-def}
C_k=\frac{g_s}{N}\,\frac{1}{\e^{\beta(\varepsilon_k-\mu)}-\eta}\equiv\frac{g_s}{N\,\eta}~\zeta_0\!
\left(\eta\lambda\e^{-\beta\varepsilon_k}\right),~~
\sum_{\bf k}C_k = 1\,.
\2
Here $\eta=+1, -1, 0$ refer to BE, FD, MB gases, respectively. $g_s\g 2s+1$ is the spin-degeneracy factor for spin $s$, $\lambda=\e^{\beta \mu}$
is the fugacity, and the function $\zeta_\nu(x)$ denotes the polylogarithm \cite{lew:81, lee:95} of order $\nu$.  The solution of the equation $\mu\equiv\mu_\eta\left(n,T \right)\g0$ gives the characteristic temperature which can
be expressed as $T_0^{(\eta)}=\varepsilon_{\rm u}/k_{\rm B} \left(6\sqrt\pi\,\zeta_{3/2}(\eta)/\eta\right)^{-2/3}$ with $\varepsilon_{\rm u}=\hbar^2k_{\rm u}^2/(2m)$ and $k_{\rm u}=2\left(6\pi^2n/g_s\right)^{1/3}$ serving as units of energy and wave number, respectively.

For further analytical discussions, Eq. (\ref{sofq}) will now be recast into an appropriate form by (i) substituting ${\bf k}\to{\bf k}+{\bf q}$ and to get $2S(q)$, (ii) expressing exponentials in $\coth$--functions in accordance with the first equation in (\ref{C_k-def}), and (iii) using $\sum_{\bf k}\left(C_k + C_{|{\bf k} +{\bf q}|}\right)=2$. The procedure finally yields  a  new form:
\1
\label{sofq-convolution}
S(q)-1=\frac{\eta N}{g_s}\sum_{\bf k}C_k\,C_{|{\bf k} + {\bf q}|}\;,~~(q>0)
\2
from which we read
\ea
\label{sum-sofq1}
\frac{1}{V}\sum_{\bf q\ne0}\e^{\ii{\bf q}\cdot{\bf r}}\left[S(q)-1\right]&=&\frac{n\eta}{g_s}\left\{
2\,C_0\sum_{\bf q\ne0}C_q\,e^{\ii{\bf q}\cdot{\bf r}} +\left|\sum_{\bf q\ne0}C_q\,\e^{\ii{\bf q}\cdot{\bf r}}\right|^2\right\}
\ee
with $C_0\g N_0(T)/N$ denoting the fraction of particles which occupy the zero--momentum state.
Introducing  the thermal de Broglie wavelength $\Lambda\g\sqrt{2\pi\hbar^2\beta/m}$ and the dimensionless function
\1
\label{F-def}
F(r)= \sum_{\bf k}\e^{\ii{\bf k}\cdot{\bf r}}C_k
=C_0(T)+\frac{2g_s}{\sqrt{\pi}\,n \Lambda^3\,\eta }\int_0^\infty\de
\kappa~\zeta_1 \!\left(\eta\lambda \e^{-\kappa^2}\right)\,\cos\left(\frac{2\sqrt{\pi}\,r}{\Lambda}\,\kappa\right)
\2
which is related to the one--body density matrix \cite{pis:03} by $n^{(1)}({\bf r},{\bf r'})\g(n/g_s)\,F(|{\bf r}-{\bf r'}|)$ , the  condensed fraction $C_0(T)$ can be extracted from the normalization condition $F(0)=1$. The evaluation of the $\kappa$--integration thus leads to
$C_0(T)=\delta_{\eta,1}\Theta(T_c-T)\left[1-\left(T/T_c\right)^{3/2}\right]$,
with $\Theta(x)$ the Heaviside unit step and  $\delta_{i,j}$ the Kronecker delta,
in conformity with \cite{bps:10a} and the condensed--IBG result \cite[Chap. 3.2]{pis:03}.

Equations (\ref{sofq-convolution}) and (\ref{sum-sofq1}) constitute to be our basic results which are valid at all temperatures and for all ideal gases.
Inserting Eq. (\ref{sum-sofq1}) into Eq. (\ref{gofrfromsofq}), we find
\1
\label{gFrelationship}
g(r)=1+\frac{\eta}{g_s}\,\left[F^2(r)-C_0^2(T)\right]
\2
which, in conjunction with  Eq. (\ref{F-def}), yields an expression in agreement with that discussed in Problem 4 of \cite[\S117]{lal:93} for an IBG at $T\!<\! T_c$. Thus Eq. (\ref{gFrelationship}),  which is valid at all temperatures for all IQGs, generalizes \cite[Eq. (117.8)]{lal:93} whose validity is for an FD gas at all $T$ but for a BE gas at $T\!>\!T_c$ only.

It seems pertinent to mention that for bosons $g(r)$ is not simply the sum of  condensate and  non--condensate (or thermal) contributions as it is for the one--body density matrix, Eq.(\ref{F-def}), or the density--response function \cite{bps:10a, pis:03}. The presence of the Fourier convolution in Eq.(\ref{sofq-convolution}) has resulted into Eq. (\ref{gFrelationship}) wherein the thermal contribution $\left[F(r)-C_0(T)\right]$ appears as a factor in the second term on the right hand side.  This factor vanishes in the limit $T\to0$ and hence is responsible for  the ``flattening'', $g(r)\to1$, observed at $T\ll T_c$.

There is another significant aspect regarding the derivation of our results  in context of the GCE used here.
On the dynamic route leading to Eq. (\ref{sofq}), and thereby Eq. (\ref{sofq-convolution}), we neither needed to nor did we use the Bogoliubov prescription which replaces $a_0$ and $a^\dagger_0$ by $c$--numbers. For a consistency check, we took recourse to the static route (not elaborated here) starting from $S(q)\g\erw{\delta N_{\bf q}\delta N^\dagger_{\bf q}}/N$ with $N_{\bf q}\g\sum_{{\bf k},\sigma}a^\dagger_{{\bf k},\sigma}a_{{\bf k}+{\bf q},\sigma}$. We obtain for a  uniform fluid the counterpart of Eq. (\ref{sofq-convolution}) wherein the right hand side contains additional terms representing correlations of number fluctuations, $\erw{\delta\left(a^\dagger_{{\bf k},\sigma}a_{{\bf k'},\sigma'}\right)~\delta\left(a^\dagger_{{\bf k'}+{\bf q},\sigma'}a_{{\bf k}+{\bf q},\sigma}\right)}$. However, for any ${\bf q}\ne0$, these extra terms vanish for  ideal gases, irrespective of population of any single--particle state. Hence it is comforting to note that  static and dynamic routes lead to exactly the same result.

\noindent  \emph{Analytic Expressions for $F(r)$}:
For a BE or an FD gas, the integral in Eq. (\ref{F-def}) can be carried
out analytically in the region $-\infty\!<\!\mu\!\le\!0$, i.e. for $0\!<\!\lambda\!\le\!1$. This region covers the complete domain of the IBG while it describes only the {\em high--T domain} ($T_0^{(-1)}\!\le\! T\!< \!\infty$) of the ideal Fermi gas (IFG).
On series expansion of $\zeta_1(z)$ and subsequent term--by--term integration,  we get
\1
\label{Fformunegative}
F(r)=C_0(T)+\frac{g_s}{n\Lambda^3\,\eta}
\sum_{\ell\g1}^\infty\frac{(\eta\lambda)^\ell}{\ell^{3/2}}\;
\exp\left(-\frac{\pi r^2}{\Lambda^2}\frac{1}{\ell}\right)\;.
\2
An alternative form, equivalent to Eq. (\ref{Fformunegative}) and most suitable for evaluation at $r\!\ll\!\Lambda$, is
\1
\label{FforsmallR1}
F(r)=C_0(T)+\frac{1-C_0(T)}{\zeta_{ 3/2}\left(\eta\lambda \right)}
\sum_{\ell\g 0}^\infty\frac{(-1)^\ell}{\ell\, !}\left(\frac{\pi r^2}{\Lambda^2}\right)^\ell\;\zeta_{\ell+3/2}(\eta\lambda),
\2
which results from series expansion of the exponential function in Eq. (\ref{Fformunegative}) and subsequent interchange of summations. Also, we have used the relation obtained on implementing $F(0)\g1$ in Eq. (\ref{Fformunegative}).

For analytical evaluation of the integral in Eq.(\ref{F-def}) for $\eta\g-1$ in the region where $\mu\!>\!0$, i.e. for $\lambda\!>\!1$,
which describes the {\em low--T domain} ($0\!\le\! T\!<\!T_0^{(-1)}$) of an IFG, we split the integral into a sum of two integrals over intervals $(0,\sqrt{\beta\mu})$ and $(\sqrt{\beta\mu},\infty)$, respectively. In the latter integral, the polylogarithm $\zeta_1(z)$ with $z\!=\!-\e^{\beta\mu-\kappa^2}$ can be expanded  into a power series in $z$, since $0\!\le \!|z|\!<\! 1$. In the former integral, where $|z|\!>\!1$, we apply the identity $\zeta_1(z)\!=\!\zeta_1(1/z)-\ln(-z)$ valid for $z\!\notin\!(0,1)$ and subsequently expand $\zeta_1(1/z)$ into a power series in $1/z$, since $\left|1/z\right|\!=\!1/|z|\!<\!1$. Term--by--term integration of the resulting infinite sum finally yields for low--$T$ IFG ($\mu\!>\!0$) :
\ea
\label{FD-Fofr-lowtemp}
F^{\rm FD}(r)&=&\frac{\tilde{k}_{\rm F}^3}{k_{\rm F}^3}\left[
3\;\frac{j_1(\tilde{k}_{\rm F}r)}{\tilde{k}_{\rm F}r}-\frac{6\pi^2}{\left(\tilde{k}_{\rm F}\Lambda\right)^3}\sum_{\ell\g1}^\infty\frac{(-1)^\ell}{\ell^{3/2}}\nonumber
\right.\\&&\left.
\hspace{-2em}\times\left\{
\frac{{\Im\,\rm erfc}\left(\frac{\sqrt{\pi}\,r}{\Lambda}\,\frac{1}{\sqrt{\ell}}-\ii \,\frac{\tilde{k}_{\rm F}\Lambda}{2\sqrt{\pi}}\,\sqrt{\ell}\right)}{\exp\left(\left(\frac{\tilde{k}_{\rm F}\Lambda}{2\sqrt{\pi}}\right)^2\,\ell-\left(\frac{\sqrt{\pi }\,r}{\Lambda }\right)^2\,\frac{1}{\ell}\right)}
+
\frac{\Re\,{\rm erfc}\left(\frac{\tilde{k}_{\rm F}\Lambda}{2\sqrt{\pi}}\,\sqrt{\ell}+\ii \frac{\sqrt{\pi}\,r}{\Lambda}\,\frac{1}{\sqrt{\ell}}\right)}{\exp\left(\left(\frac{\sqrt{\pi }\,r}{\Lambda }\right)^2\,\frac{1}{\ell}-\left(\frac{\tilde{k}_{\rm F}\Lambda}{2\sqrt{\pi}}\right)^2\,\ell\right)}
\right\}\right]
\ee
where $j_1(x)$ is the spherical Bessel function of the first kind and order 1, and ${\rm erfc}(x)$ is the complementary error function. Also, $\tilde{k}_{\rm F}\!=\!\hbar^{-1}\sqrt{2m\mu_{-1}(n,T)}$ denotes a generalized Fermi wave number with $\lim_{T\to0}\tilde{k}_{\rm F}\!=\!k_{\rm F}\!=\!(6\pi^2 n/g_s)^{1/3}$ and is a measure of the chemical potential.
In the low--temperature limit, i.e. for $T\!\to\!0$ and $\Lambda\!\to\!\infty$, one easily retrieves the following from Eq. (\ref{FD-Fofr-lowtemp}): $F^{\rm FD}(r)~\stackrel{T\to0}{\longrightarrow}~3\,j_1(k_{\rm F} r)/(k_{\rm F}r)$, the expression given, see e.g., in Ref. \cite{lel:95}.

\noindent \emph{Pair Distribution Functions}:
From Eq. (\ref{gFrelationship}), one readily finds for a  dilute quantum gas, $g^{\rm MB}(r)\!=\!1$  which coincides with the  classical ideal--gas result.
One also deduces the results
$g(0)\!=\!1+(\eta/g_s)[1-C_0^2(T)]$ and $g(\infty)\!=\!1$ leading to the following bounds:\\
$g^{FD}(0)\! \equiv\!  1-1/g_s \!\le\! g^{FD}(r) \!\le\! g^{FD}(\infty)\! \equiv\!  1$ and
$g^{BE}(\infty) \!\equiv\!  1\! \le\!  g^{BE}(r)\! \le \! g^{BE}(0) $ with
\1
\label{pile-height}
g^{BE}(0)=\left\{\begin{array}{ccc}
1+ 1/g_s &\mbox{if}&T> T_c\\
1+ (1/g_s)\left[2\left(T/T_c\right)^{3/2}-\left(T/T_c\right)^{3}\right]&\mbox{if}&T\le T_c
\end{array}
\right.\;.
\2
The small-- and large--$r$ asymptotic behaviors for $T\!<\!T_c$ are obtained as
$g^{BE}(r\!\to\! 0)\!=\!g^{BE}(0)-\left(2\pi r^2/g_s \Lambda^2\right)\left(T/T_c\right)^{3/2}\left[\zeta(5/2)/\zeta(3/2)\right]
+ {\cal O}\left((r/\Lambda)^4\right)$ and
$g^{BE}(r\!\to\!\infty)\!=\!1+ 2 C_0 (T)/\left(n\Lambda^2\,r\right)+ g_s/\left(n^2\Lambda^4\,r^2\right)$.
We find that the latter asymptote improves the expression given in \cite[Eq.(21)]{bae:71}, and is in agreement with \cite[p.359]{lal:93}; the expression given in \cite{bae:71} would be valid only at $T\!\ll\! T_c$ whereas the validity of ours is in the entire range $0\!\le\! T\!\le\! T_c$.
Also, for $T\!\to\! 0$, one gets
$g^{FD}(r\!\to\! 0)\!=\!
(g_s-1)/g_s +\left(k_F^2 r^2/5\,g_s\right)\left[1- 3k_F^2 r^2/35 +{\cal O}(r^4)\right]$
wherein the first two terms on the right--hand side give the
result as obtained by Lee and Long \cite{lel:95} while discussing the
static structure for an ideal electron gas at $T\g 0$. And for large distance $r$, we get  $g^{FD}(r\!\to\! \infty)\!=\!1 - 9\left[\cos^2(k_{\rm F} r)- \sin(2k_{\rm F} r)/(k_{\rm F}r)\right]/(g_s k_{\rm F}^4 r^4)+{\cal O}(r^{-6})$ approaching unity as $r^{-4}$ by damped oscillations,  which improves on a  result given in \cite[\S117]{lal:93}.

On substituting in Eq. (\ref{gFrelationship}) the high--$T$ asymptote of $F(r)$ deduced from Eq.(\ref{Fformunegative}), the Gaussian form,
\1
\label{gvonr-high-T-asymp}
g(r)~\stackrel{T\gg
T_0^{(\eta)}}{\longrightarrow}~
1+\frac{\eta}{g_s}\,\exp\left(-\frac{2\pi\,r^2}{\Lambda^2}\right)
\2
is obtained. In fact, the asymptote (\ref{gvonr-high-T-asymp}) generalizes to the spinor gases the earlier results derived by others for zero-spin particles, see e.g., \cite{pat:96}.  Although the experimental results for a thermal bosonic gas were fitted by Schellekens \emph{et al.} \cite{sch:05} using an expression like Eq. (\ref{gvonr-high-T-asymp}),  a detailed analysis together with the theoretical plots showing contrasting behavior for spinor bosons and fermions is lacking.

\begin{figure}\begin{center}
\includegraphics[width=100mm,angle=0]{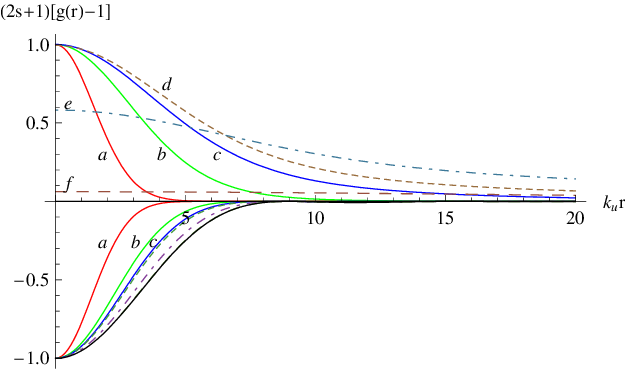}
\parbox{130mm}{\caption{\label{FNIP-pair-corr}\small
(Color online). Pair correlation functions of  ideal BE (above abscissa), FD (below abscissa), and
MB (coinciding with the abscissa for all $T$) gases at reduced temperatures $T^*\g$ 0.491 (a, red), 0.164 (b, green), 0.114 (c, blue), 0.104 (d, short--dashed), 0.055 (e, dot--dashed), and 0.011 (f, long--dashed).  $T\g0$ results: bosons (coinciding with abscissa),
fermions (lowermost curve, black, shadowing the long--dashed line).
}}\end{center}
\end{figure}

The  computed values of $(2s+1)\left[g(r)-1\right]=\eta\left[ F^2(r)-C_0^2(T)\right]$ are depicted in
Fig. \ref{FNIP-pair-corr} for a set of six reduced temperatures $T^*\equiv k_B T/\varepsilon_u \g 0.491, ~0.164, ~0.114, 0.104, ~0.055$ and $0.011$
corresponding to $T/T_c\g4.5,~1.5,~1.05,~0.95,~0.5,~0.1$, respectively. The value $g^{MB}(r)\g 1$ of an  ideal dilute quantum gas coincides with the abscissa at all $T$ and, it is to be noted, that the pair--correlation properties of an IQG differ qualitatively from those of the corresponding dilute gas, even at high temperatures. As displayed in Fig.\ref{FNIP-pair-corr}, $g^{\rm BE}(r)\!\ge\!1$ and $g^{\rm FD}(r)\!\le\!1$ at all $T$.
We term the appearance of a bump (dip) as the ``Bose pile" (``Fermi hole") which reflects the statistical ``effective" attraction (repulsion) in ideal gases, an interaction which weakens with increasing spin.  For $s\!\gg\!1$, one would recover the result $g(r)\g1$ true for a classical  ideal gas.

It is found in low--$r$ region that $g^{FD}(r)$ monotonically decreases (increases in magnitude) with increase in $r$ as $T$ decreases. However, for bosons in that domain, $g^{BE}(r)$ increases as $T$ decreases for $T>T_c$ and the trend reverses with decrease in $T$ below $T_c$.  Let us look further into the behavior of fermionic curves at all $T$ and bosonic at $T> T_c$, i.e. those marked (a) to (c) above abscissa. It can be seen that the pile (hole) being generated by rotating
the Gaussian curves about the ordinate merely narrows down retaining its original
spin--dependent height (depth) as T is increased. They ultimately acquire congruent bell-shaped forms (cf. Eq.(\ref{gvonr-high-T-asymp})) in the high--$T$ regime having width (at
half maximum) $\g\sqrt{\ln 2/2\pi}~\Lambda\approx0.332\,\Lambda$. Also, it can be seen that $g^{BE}(r)$ and $g^{FD}(r)$ show different behaviors at $T=4.5\,T_c$ as opposed to studies wherein it was found that response functions
\cite{bps:10a}, dynamical structure factors \cite{bos:08}, and
momentum distribution functions \cite{bps:10a} of BE, FD, and MB
gases are essentially independent of statistics at this temperature. However, in the limit $T\!\to\!\infty$, we get $g(r)\g1$ {\em independent}
of statistics, as expected.

The bosonic curve marked (c) clearly demonstrates that
$g^{BE}(r)$ becomes  long--ranged as $T$ reaches in the close vicinity of  $T_c$ from high--$T$ side. However,
unlike this,  $g^{FD}(r)$ is of much
shorter range at all $T$.   It is estimated that the  correlation length  $\xi_{\rm
FD}(T)\le10\,k_{\rm
u}^{-1}$ which presents a measure of the largest distance at which
fermion pairs are still correlated. If $T$ is raised, $\xi_{\rm
FD}(T)$ further decreases and at high $T$ it
is of the order of $\Lambda$ (cf. Eq.(\ref{gvonr-high-T-asymp})).

It is tempting to compare both the curves marked (a) in Fig. \ref{FNIP-pair-corr}
with Fig. 2 of Ref. \cite{jel:07} wherein the HBT effect showing bunching and antibunching for ultracold atoms of $^4$He$^*$ and  $^3$He$^*$ have been depicted. The striking resemblance between the theoretical and experimental plots is remarkable. The direct comparison of our curves with the experimental ones will have to await availability of data free from the uncertainty and systematic errors in measurements,  mentioned  by Jeltes \emph{et al.} \cite{jel:07}.  However, the interesting physics is unfolded in the BEC phase represented by the curves  (d) to (f). The height of the Bose pile goes on decreasing  and the curve gets increasingly flattened as $T\to0$ (cf. Eq. (\ref{pile-height})). The curve (f) for $T\!=\!0.1 T_c$ is almost flat and ultimately the plot for an IBG at $T\!=\!0$ coincides with the abscissa analogous to the Bose-condensed phase experimental result \cite{sch:05, g(r):expt} revealing that the system is completely coherent. In fact, it is just like the situation  in a single-mode laser in which the photons are not bunched \cite{photons}. Furthermore, our studies depict temperature-dependent aspects of bunching and antibunching.

\noindent \emph{Acknowledgments}:
The work is partially supported by the Indo--German (DST--DFG) collaborative research program. JB and KNP gratefully acknowledge financial support from the Alexander von Humboldt Foundation.

\noindent *Address for correspondence. Email: gss.phy@iitr.ernet.in


\begin{thebibliography}{10}


\bibitem{hat:56}
R.~Hanbury~Brown and R.~Q. Twiss, Nature (London) \textbf{177}, 27 (1956).

\bibitem{san:10}
C. Sanner \emph{et al.}, Phys. Rev. Lett. \textbf{105}, 040402 (2010).

\bibitem{mue:10} T. Mueller \emph{et al.}, Phys. Rev. Lett. \textbf{105}, 040401 (2010).

\bibitem{jel:07}
T. Jeltes \emph{et al.}, Nature (London) \textbf{445}, 402 (2007).

\bibitem{sch:05}
M. Schellekens \emph{et al.}, Science \textbf{310}, 648 (2005).

\bibitem{lal:93}
L.~D. Landau and E.~M. Lifshitz.
\newblock{\em Statistical Physics}, \emph{Part I}, 3rd ed.
\newblock(Pergamon Press, Oxford, UK, 1993).

\bibitem{lur:68}
M.~Luban and M.~Revzen, J. Math. Phys. \textbf{9}, 347 (1968).

\bibitem{gub:68}
J.~D. Gunton and M.~J. Buckingham, Phys. Rev. \textbf{166}, 152 (1968).

\bibitem{bae:71}
K.~Baerwinkel, Phys. Kondens. Materie \textbf{12}, 287 (1971).

\bibitem{lel:95}
M.~H. Lee and M.~Long, Phys. Rev. E \textbf{52}, 189 (1995).

\bibitem{gsb:00}
P.~Gori-Giorgi, F.~Sacchetti, and G.~B. Bachelet, Phys. Rev. B \textbf{61}, 7353 (2000).

\bibitem{lew:81}
L.~Lewin.
\newblock {\em Dilogarithms and Associated Functions},
\newblock (Elsevier, New York, 1981).

\bibitem{lee:95}
M.~H. Lee, J. Math. Phys. \textbf{36}, 1217 (1995).

\bibitem{g(r):11}
The polylogarithm has proved useful earlier for IQG in providing a unified
description of statistical thermodynamics in \cite{lee:95, lek-lee},
and also the dynamics in \cite{bps:10b, bps:10a, bos:08}.

\bibitem{lek-lee}
M.~H. Lee and K.~Jim, Physica A \textbf{304}, 421 (2002); M.~H. Lee, Acta Physica Polonica B \textbf{40}, 1279 (2009).

\bibitem{bps:10b}
J.~Bosse, K.~N. Pathak, and G.~S. Singh, Physica A \textbf{389}, 1173 (2010).

\bibitem{bps:10a}
J.~Bosse, K.~N. Pathak, and G.~S. Singh, Physica A \textbf{389}, 408 (2010).

\bibitem{nag:99}
M.~Naraschewski and R.~J.~Glauber, Phys. Rev. A \textbf{59}, 4595 (1999).

\bibitem{pol:96}
H.~D.~Politzer, Phys. Rev. A \textbf{54}, 5048 (1996).

\bibitem{gom:06}
J.~Viana~Gomes \emph{et el.}, Phys. Rev. A \textbf{74}, 053607 (2006).

\bibitem{pis:03}
L.~Pitaevskii and S.~Stringari.
\newblock {\em Bose--Einstein Condensation}.
\newblock (Clarendon Press, Oxford, 2003).

\bibitem{pat:96}
R.~K. Pathria, {\em Statistical Mechanics},
\newblock {2nd ed. (Butterworth-Heinemann, Oxford, 1996)}.

\bibitem{bos:08}
J.~Bosse and T.~Schlieter.
\newblock in {\em Proceedings of the 9th
Intnternational Conference on Path Integrals -- New Trends and Perspectives, 2007,}
\newblock {edited by W.~Janke and A.~Pelster (World Scientific, Singapore, 2008), pp. 409-412.}

\bibitem{g(r):expt}
Although the BEC phase result in Fig. 2 of Ref. \cite{sch:05} is not at $T\ll T_c$
but the fitted curve therein is flat probably due to less accuracy in measurements.

\bibitem{photons}
R.~J. Glauber, Phys. Rev. Lett. \textbf{10}, 84 (1963); F.~T. Arecchi, E.~Gatti, and A. Sona, Phys. Lett. \textbf{20}, 27 (1966).

\end{thebibliography}
\end{document}